\documentclass[aps,prl,twocolumn,showpacs,preprintnumbers,
amsmath,amssymb]{revtex4}
\usepackage{graphicx}
\usepackage{bm}

\begin{document}

\title{Aharonov-Casher effect in Bi$_{\rm
2}$Se$_{\rm 3}$ square-ring interferometers}

\author{Fanming Qu}
\author{Fan Yang}
\author{Jun Chen}
\author{Jie Shen}
\author{Yue Ding}
\author{Jiangbo Lu}
\author{Yuanjun Song}
\author{Huaixin Yang}
\author{Guangtong Liu}
\author{Jie Fan}
\author{Yongqing Li}
\author{Zhongqing Ji}
\author{Changli Yang}
\author{Li Lu}
\email[Corresponding authors: ]{lilu@iphy.ac.cn}

\affiliation{Daniel Chee Tsui Laboratory, Beijing National
Laboratory for Condensed Matter Physics, Institute of Physics,
Chinese Academy of Sciences, Beijing 100190, People's Republic of
China}

\date{\today}% It is always \today, today,

\begin{abstract}
Electrical control of spin dynamics in Bi$_{\rm 2}$Se$_{\rm 3}$ was
investigated in ring-type interferometers. Aharonov-Bohm and
Altshuler-Aronov-Spivak resistance oscillations against magnetic
field, and Aharorov-Casher resistance oscillations against gate
voltage were observed in the presence of a Berry phase of $\pi$. A
very large tunability of spin precession angle by gate voltage has
been obtained, indicating that Bi$_{\rm 2}$Se$_{\rm 3}$-related
materials with strong spin-orbit coupling are promising candidates
for constructing novel spintronic devices.
\end{abstract}

% PACS, the Physics and Astronomy Classification Scheme.

%\keywords{topological insulators, nanowires, Aharonov-Casher effect}

\maketitle

One important task of spintronics research is to explore electrical
control of spin dynamics in solid-state devices via spin-orbit
coupling (SOC) \cite{1}. By tuning applied gate voltage one can
rotate the spin of a moving electron with purely electrical means
rather than traditionally with magnetic fields. The devices ever
proposed and studied include the Datta-Das spin field effect
transistors based on Rashba SOC \cite{2,3}, the spin interferometers
and filters \cite{4,6,5,7,8,9} via Aharonov-Casher (AC) effect
\cite{10}, and the quantum-dot spin qubits using electrical pulse to
control the spin precession \cite{11,12,13}, etc. Since the
tunability of spin rotation by applied gate voltage relies on the
strength of SOC, materials with stronger SOC would preferably be
chosen to construct devices in this regard.

Topological insulators (TIs) \cite{14,15,16} are a new class of
materials with strong SOC. The SOC in TIs is so strong that it leads
to the formation of helical electron states at the surface/edge with
inter-locked momentum and spin degrees of freedom. Many novel
properties of TIs have been observed, including the formation of
Dirac fermions \cite{17,18,19,20}, suppression of backscattering
\cite{21}, and the appearance of weak anti-localization (WAL)
\cite{22,23,24,25,26}.

With such a novel electron system caused by strong SOC, electrical
control of spin dynamics in TIs becomes a particularly interesting
issue. Previously, Molenkamp group has studied the spin interference
in a ring device made of HgTe/HgCdTe quantum well, a material which
could be tuned into a two-dimensional (2D) TI, and reached a
tunability higher than that in devices based on InAlAs/InGaAs two
dimensional electron gas (2DEG) \cite{5,9}. Here we report our
investigation on the electrical control of spin dynamics in a
square-ring type of interferometer based on Bi$_{\rm 2}$Se$_{\rm
3}$, a material which could eventually be tuned into a 3D TI. The
tunability of spin precession by gate voltage is found to be
significantly higher than that reported before.

Bi$_{\rm 2}$Se$_{\rm 3}$ nanoplates were synthesized in a 2-inch
horizontal tube furnace via a chemical vapor deposition method
similar to literature \cite{29}. Figure 1a and 1b are the scanning
electron microscopy image and the high-resolution transmission
electron microscopy image of the single-crystalline Bi$_{\rm
2}$Se$_{\rm 3}$ nanoplates, respectively. Figure 1c indicates that
the nanoplates were grown along [11${\overline 2}$0] direction.
Figure 1e is the X-ray powder diffraction pattern of the nanoplates,
confirming they are of Bi$_{\rm 2}$Se$_{\rm 3}$ phase. The
nanoplates of several tens nm thick were transferred to highly doped
Si substrates with 300 nm SiO$_2$. E-beam lithography and reactive
ion etching with Ar gas at low power (20 W) were used to pattern the
devices. The real interferometer devices used in this experiment
contained two parallel rows of square rings, with four rings in each
row, as shown in Fig. 1d (note that the whitish colloid is residual
PMMA mask). Electron transport measurements were carried out in a
sorption-pumped $^3$He cryostat with standard lock-in technique,
after Pd electrodes were fabricated to the two ends of the rows.

The data presented here were obtained from two devices with the same
structure but slightly different thickness, 17 nm and 11 nm, labeled
as device \#1 and \#2, respectively. For device \#1, we were able to
pattern a Hall bar on the same nanoplate for material
characterization. Figure 1f shows the magnetic field dependencies of
the longitudinal resistance $R_{\rm xx}$ and Hall resistance $R_{\rm
xy}$ of the nanoplate at 0.3 K. Deduced from the Hall slope, $R_{\rm
H}=-1/ne$, the areal electron density of the nanoplate is
$n=3.9\times 10^{13}$ cm$^{-2}$. Assuming a simple Drude form
$\sigma=ne\mu$, the effective electron mobility is estimated to be
$\mu$=1000 cm$^2$/Vs.

\begin{figure}
\includegraphics[width=1.0 \linewidth]{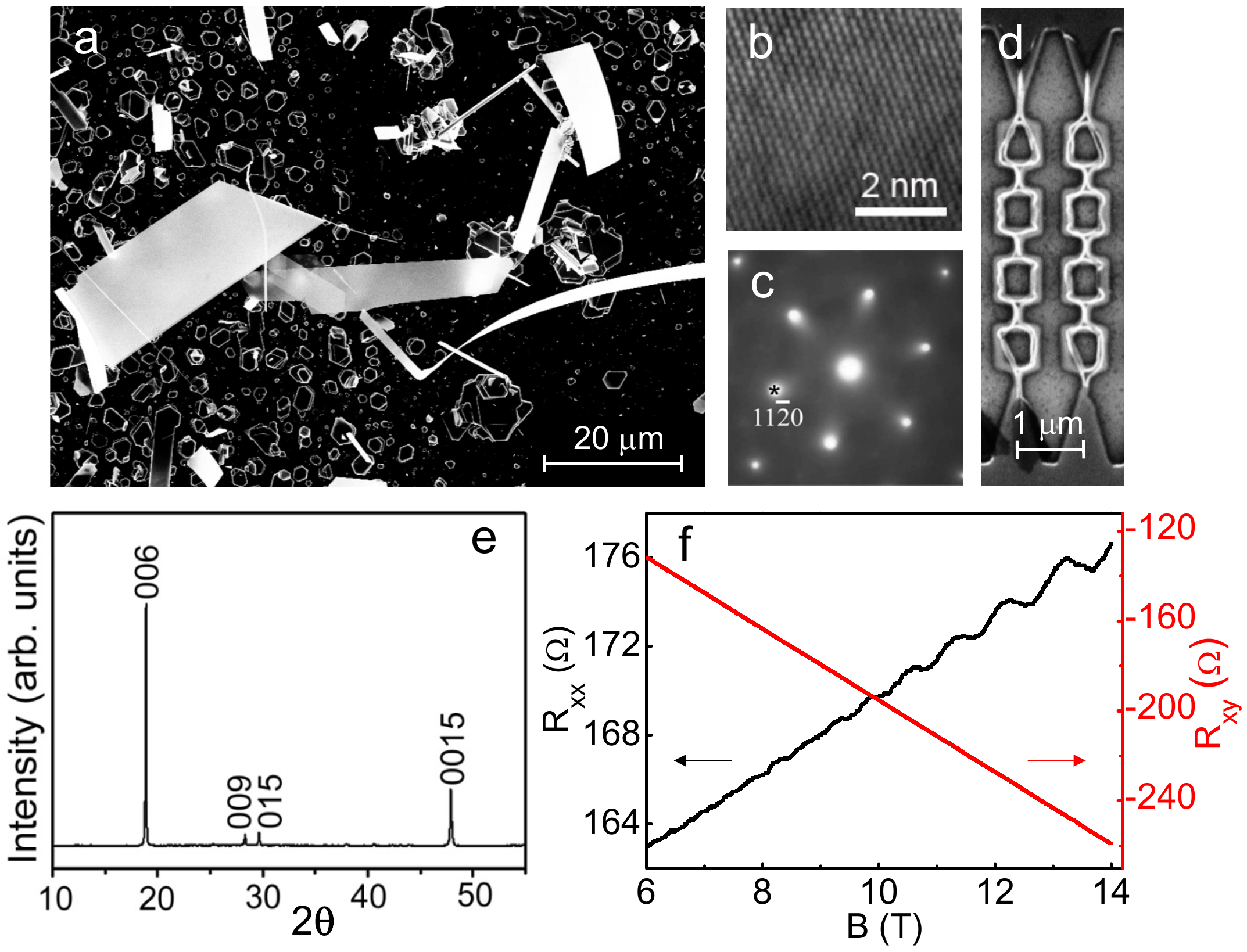}
\caption{\label{fig:FMQ_Fig1} {(color online) (a) Scanning electron
microscopy (SEM) image of the as-grown single-crystalline Bi$_{\rm
2}$Se$_{\rm 3}$ nanoplates and nanoribbons on Si substrate. (b)
High-resolution transmission electron microscopy image of a
nanoplate. (c) Electron diffraction pattern of a nanoplate with
growth direction along [11${\overline 2}$0]. (d) SEM image of the
square ring devices. The whitish colloid on top of the rings is the
residual PMMA mask. (e) X-ray powder diffraction pattern consistent
with PDF cards No. 33-214 confirms the Bi$_{\rm 2}$Se$_{\rm 3}$
phase. (f) $R_{\rm xx}$ and $R_{\rm xy}$ curves of a Hall-bar
fabricated on the same nanoplate connecting to device \#1. The data
were taken at 0.3 K. }}
\end{figure}

Figure 2a shows the magneto-resistance (MR) of device \#1 measured
at 0.3 K. A sharp dip appears around zero magnetic field, caused by
WAL of the electrons due to the existence of a Berry phase of $\pi$
in the material \cite{22,23,24,25,26}. The MR away from B=0
demonstrates typical features of the universal conductance
fluctuations (UCF), with a rough period of $\sim$0.1 to 0.2 T, and a
decaying amplitude with increasing both the magnetic field $B$ and
temperature $T$ (see the inset of Fig. 2a). On the UCF background,
smaller but reproducible resistance oscillations at a period of
$\sim$0.013 T can be recognized up to the highest field of this
experiment, 8T. Figure 2b and 2c show the details of oscillations in
the two marked windows in Fig. 2a at low and high magnetic fields,
respectively.

\begin{figure}
\includegraphics[width=0.85 \linewidth]{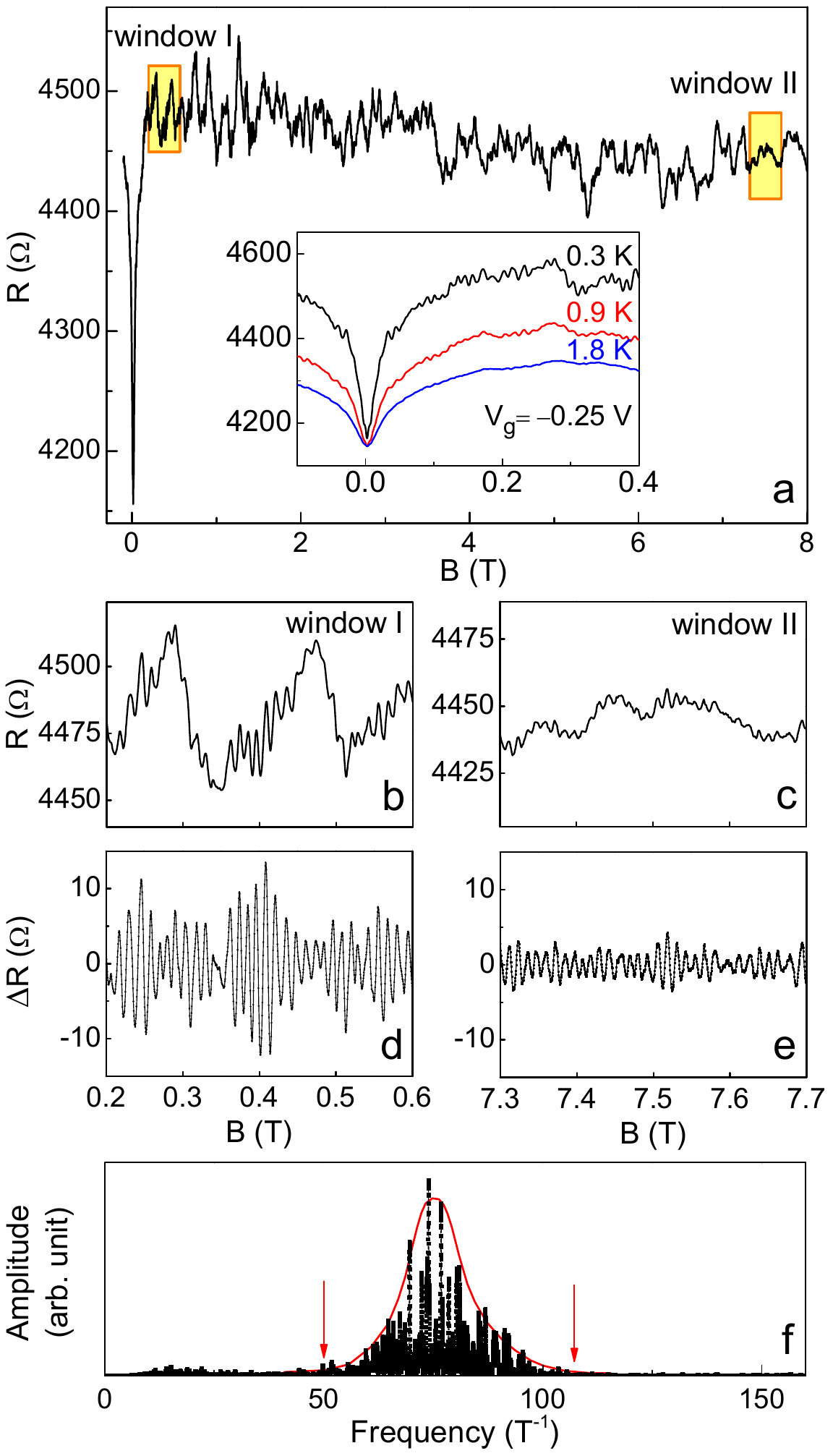}
\caption{\label{fig:FMQ_Fig2} {(color online) (a) Magneto-resistance
(MR) of device \#1 measured at 0.3 K and $V_{\rm g}$=0. Inset: MR
around zero field and at several different temperatures. (b) and
(c): Details of MR fluctuations in low and high magnetic fields, at
the two windows marked in (a), respectively. (d) and (e): MR in (b)
and (c) after the UCF background is subtracted, labeled as $\Delta
R$. (f) Fast Fourier transformation (FFT) of $\Delta R$ in a field
range of 0.2 to 8 T. The red line and arrows are guidance for the
eyes, showing the peak position and the upper and lower bounds
expected from the inner-most, middle and out-most electron
trajectories of the square rings.}}
\end{figure}

The MR oscillations become much clearer after the UCF background is
subtracted, as shown in Fig. 2d and e. To do so, we let each data
point subtract its moving averaged surroundings over $\sim$ 0.1 T
span. It can be seen that the oscillations of $\Delta R$ sustain up
to the highest field of this experiment, although the amplitude
decreases significantly. Fast Fourier transformation (FFT) analysis
of the data from 0.2 to 8 T reveals a broad peak centered at
$\sim$76 T$^{-1}$ (Fig. 2f). It is in perfect agreement with the
expected frequency of Aharonov-Bohm (AB) oscillations $(\Delta
B)^{-1}=S/\phi_0=$75 T$^{-1}$, where $\phi_0=h/e$ is the flux
quanta, $h$ the Plank's constants, $e$ the electron charge, and
$S=$560 nm$\times$560 nm is the averaged area of the rings. The
lower and upper bounds of the FFT peak, as marked by the red arrows
in the figure, also correspond nicely to the areas of the outer-most
and inner-most trajectories of electrons in the rings whose width is
$\sim$120 nm.

Figure 3a and b are 2D plots of $\Delta R$ as a function of magnetic
field $B$ and gate voltage $V_{\rm g}$ for devices \#1 and \#2,
respectively. In these plots, the UCF and the predominant WAL dip at
each row of the data are largely subtracted by the above mentioned
treatment. Robust AB periods extending to higher fields can be
clearly seen, as illustrated by the long arrows. Besides, a second
type of periods with doubled frequency can also be recognized at
fields below $\sim$0.01 T, as indicated by the short arrows. These
are the Altshuler-Aronov-Spivak (AAS) oscillations arisen from the
interference of electron's wave packet circulating along the two
time-reversal loops on the ring. AAS oscillation is a low field
phenomenon, because the magnetic field breaks the time-reversal
symmetry of the two trajectories. However, the fast decay of AAS
oscillations here is largely caused by the finite width of the
rings. As shown in Fig. 2f, the expected frequencies of oscillations
differ by a factor of $\sim$2 between the inner-most and out-most
trajectories on the rings. It smears out the oscillations at high
fields.

\begin{figure}
\includegraphics[width=1.0 \linewidth]{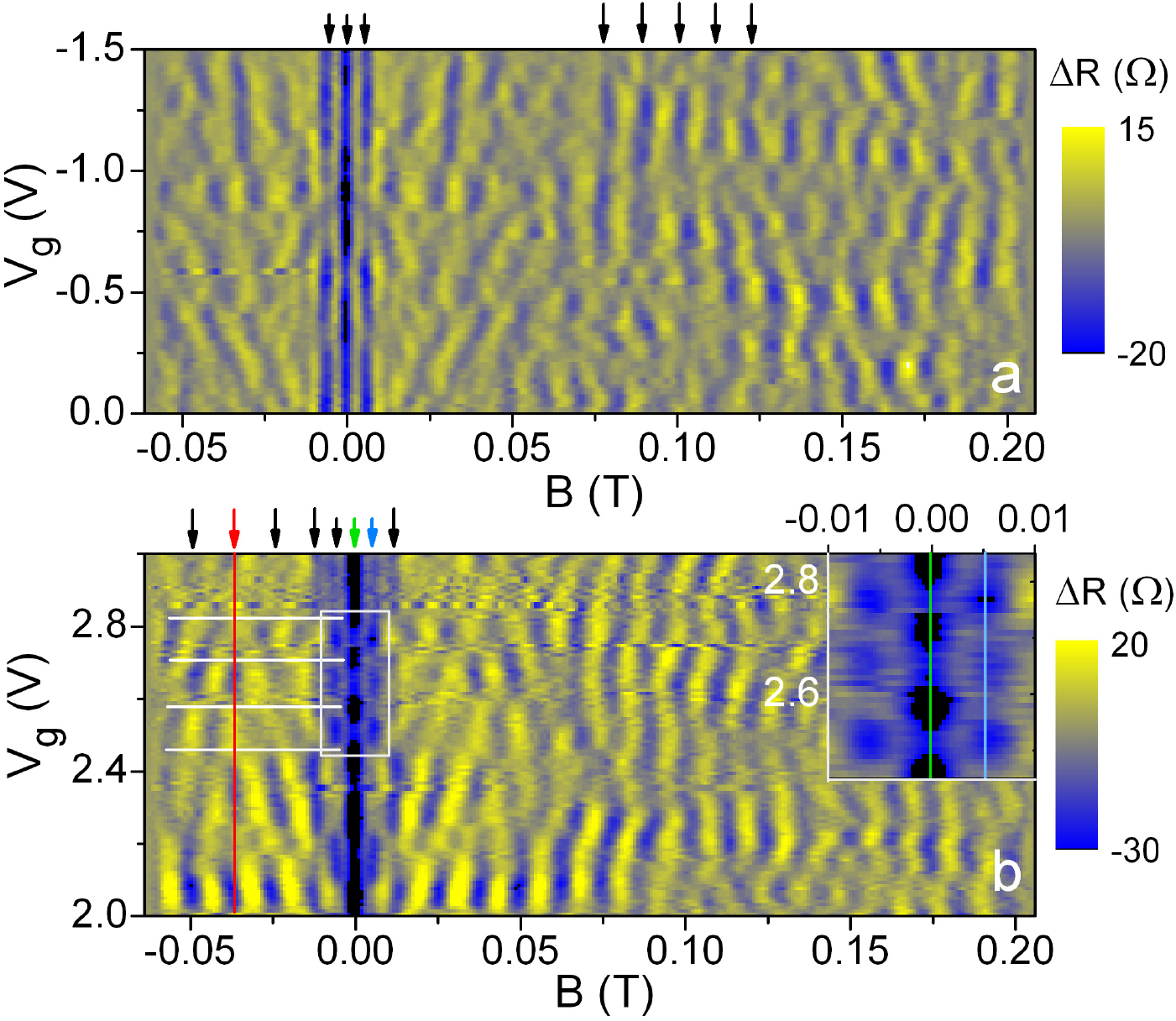}
\caption{\label{fig:FMQ_Fig3} {(color online) 2D plots of $\Delta R$
as a function of magnetic field $B$ and back-gate voltage $V_{\rm
g}$ measured at 0.3 K for devices \#1 (a) and \#2 (b). The long and
short arrows in both plots illustrate the $B$-dependent AB and AAS
oscillation periods, respectively. The white lines in (b) indicate
the $\pi$ phase shift of the $V_{\rm g}$-dependent AC oscillations
in the AB region. The $V_{\rm g}$-dependent oscillation frequency
doubles in the AAS region. The rectangle area in (b) is magnified
and shown at the up-right corner. }}
\end{figure}

Besides the AB and AAS oscillations, one important feature
demonstrated in Fig. 3 is the resistance fluctuations with varying
$V_{\rm g}$. Relatively regular oscillations can be recognized in
Fig. 3b at fields below 0.05 T. The places where $\pi$ phase shift
happens are marked by the white lines in the AB oscillation region.
In the AAS oscillation region where the period in $B$ halves, the
period in $V_{\rm g}$ also halves.

\begin{figure}
\includegraphics[width=0.95 \linewidth]{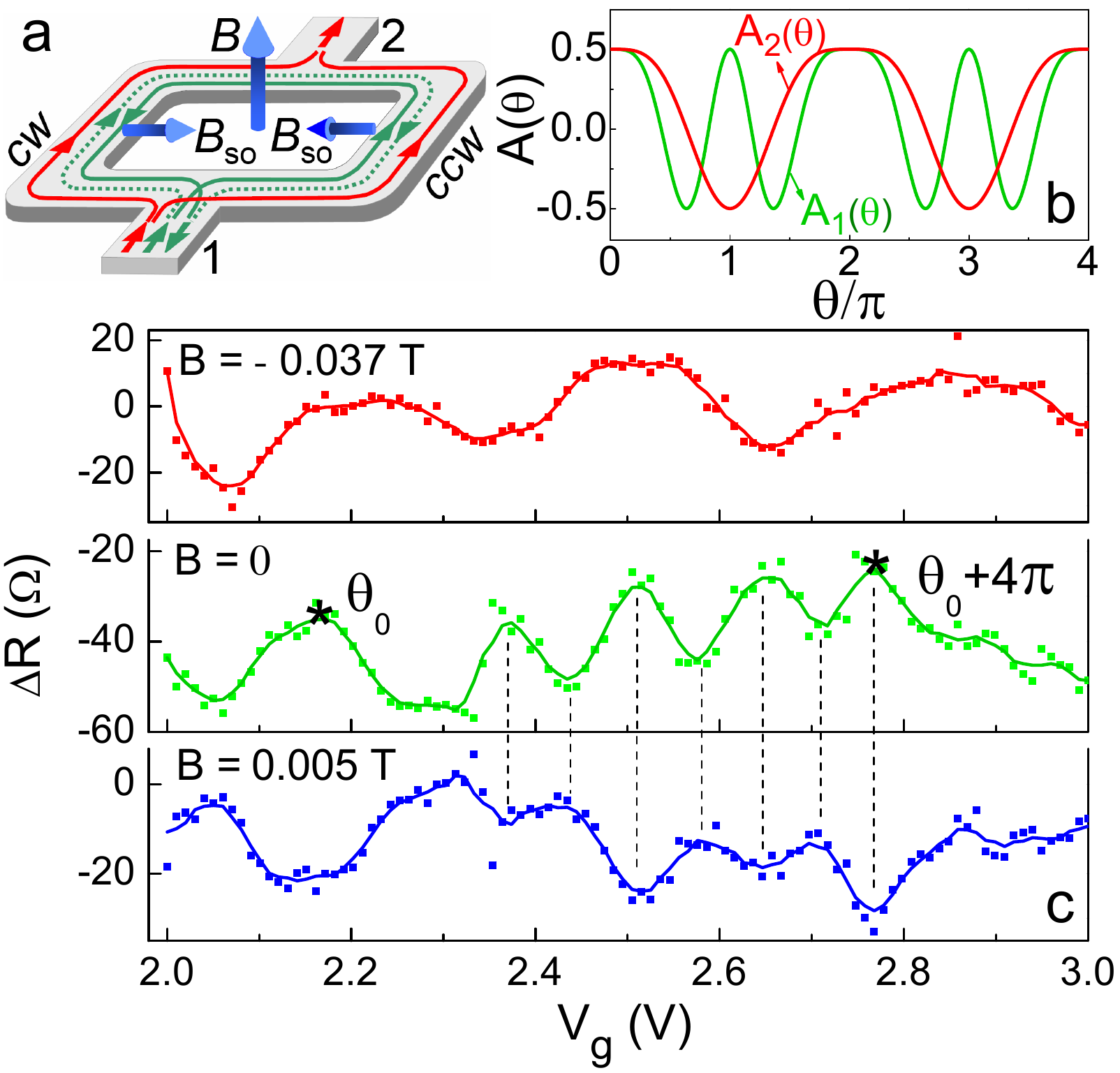}
\caption{\label{fig:FMQ_Fig4} {(color online) (a) Illustration of AB
interference (red trajectories) and AAS interference (solid and
dottd green loops) for charges in a square ring. The existence of
SOC creates an effective magnetic field $B_{\rm so}$ pointing
towards/from the center of the ring for counterclockwise
(CCW)/clockwise (CW) propagation modes, which influences spin
precession and generates an AC phase in addition to the AB and AAS
phases. (b) $A_1(\theta)$ and $A_2(\theta)$ as a function of
$\theta$ (see text). The former varies roughly at twice frequency of
the latter. (c) $\Delta R-V_{\rm g}$ curves taken from Fig. 3b at
fixed fields marked by the lines and arrows of corresponding colors.
The spin precession angle is modulated by 4$\pi$ by varying $V_{\rm
g}$ from the interval 2.16 to 2.77 V marked by the two stars. The
dashed lines help illustrating the opposite phases between the green
and blue curves in the AAS region.}}
\end{figure}

$V_{\rm g}$-dependent resistance oscillations have previously been
studied in circle rings \cite{4,5,9,30,31} and square rings
\cite{6,7,8} based on 2DEG with SOC. The oscillations were
attributed to $V_{\rm g}$-modulated spin interference via SOC, and
were referred to as an AC effect.

It is known that a charged particle circulating around a magnetic
flux acquires an AB or AAS phase. As an electromagnetic dual, a
magnetic moment circulating around an electric flux will similarly
acquire an AC phase. It has been pointed out \cite{32,33} that AB
and AC phase can be understood in a unified picture by regarding
Rashba and Dresselhaus SOC in two dimensions as a Yang-Mills
non-Abelian gauge field, whose non-commutativity generates a phase
difference (AC phase) between the two paths in a square ring.

For a more detailed analysis of the data, let us consider a 1D
square ring illustrated in Fig. 4a. The SOC induces an effective
magnetic field $B_{\rm so}$ perpendicular to the momentum and the
electric field. For an electron of given spin orientation injected
from terminal 1 and propagating along the AB or AAS trajectories,
the energy dispersion relations for the cases of spin
antiparallel/parallel to $B_{\rm so}$ are
$E=\hbar^2k^2/2m^*\pm\alpha k$, where $m^*$ is the effective mass of
electron and $\alpha$ is the Rashba SOC parameter. The spin
precession angle \cite{2} over a distance $L$ along a straight
channel is $\theta =\Delta kL=2\alpha m^*L/\hbar^2$, here $L$ is the
side length of the square. When two partial waves propagate around
the ring, the spin precession axis ($B_{\rm so}$) will change
directions from side to side, and the orders are different for
clockwise (CW) and counterclockwise (CCW) modes, which results in an
additional phase (AC phase) besides the AB/AAS phase when the two
partial waves interfere at terminal 2/1, as schematically shown in
Fig. 4 (a). The probability of finding an electron at terminal 1 is
\cite{6}:
\begin{equation}
\begin{aligned}
\overline{\langle
\Psi_1|\Psi_1\rangle}&=\frac{1}{2}+\frac{1}{4}(\cos^4\theta
+4\cos\theta \sin^2\theta +\cos 2\theta)\cos\phi_1\\
&\equiv\frac{1}{2}+A_1(\theta )\cos\phi_1
\end{aligned}
\end{equation}
\noindent where $\phi_1=2eBL^2/h$ is the AAS phase.

On the other hand, the probability of finding an electron at
terminal 2 is \cite{8}:
\begin{equation}
\begin{aligned}
\overline{\langle
\Psi_2|\Psi_2\rangle}&=\frac{1}{2}+\frac{1}{4}(\sin^2\theta +\cos 2\theta)\cos\phi_2\\
&\equiv\frac{1}{2}+A_2(\theta )\cos\phi_2
\end{aligned}
\end{equation}
\noindent where $\phi_2=eBL^2/h$ is the AB phase.

From the two equations above, the amplitude of AAS and AB
oscillations will be modulated by $A_1(\theta )$ and $A_2(\theta )$,
respectively. The $\theta$ (thus $V_{\rm g}$) dependence of
$A_1(\theta )$ and $A_2(\theta )$ are plotted in Fig. 4b. One can
see that $A_1(\theta )$, which is associated with the AAS
trajectories, fluctuates roughly at twice the frequency of
$A_2(\theta )$ which is associated with the AB trajectories. This is
in agreement with our data. As can be seen in Fig. 4c, among the
three $\Delta R-V_{\rm g}$ curves picked up in Fig. 3b at three
fixed fields (marked by lines and arrows of corresponding colors),
the green and blue ones picked up from the AAS region oscillate
roughly at a doubled frequency compared to the red one which is
picked up from the AB region.

The tunability of $V_{\rm g}$ can be estimated as follows. By
comparing the data in Fig. 4c to the curves in 4b, the spin
precession angle is modulated by 4$\pi$ between a $V_{\rm g}$ span
of 2.16 to 2.77 V (marked by the two stars), so that
$\Delta\theta/\Delta V_{\rm g}=6.6\pi$/V. Since $\theta=2\alpha
m^*L/\hbar^2$, the tunability of $V_{\rm g}$ on $\alpha$ is
$\Delta\alpha/\Delta V_{\rm g}=(\hbar^2/2m^*)\Delta\theta/L\Delta
V_{\rm g}$=11 (peVm)/V (where $m^*=0.13 m_e$ for Bi$_{\rm
2}$Se$_{\rm 3}$ \cite{34}). Thus, the tunability of our Bi$_{\rm
2}$Se$_{\rm 3}$ devices is an order of magnitude larger than that of
InAlAs/InGaAs devices \cite{5,7}, and is estimated to be more than
two times larger than that of HgTe/HgCdTe devices \cite{9}. If
taking the thickness of the insulating layer into account (e.g., 300
nm for our Bi$_{\rm 2}$Se$_{\rm 3}$ devices, 50 to 100 nm for
InAlAs/InGaAs devices, and 100 to 200 nm for HgTe/HgCdTe devices),
the tunability here is even higher.

It has to be noted that with an areal carrier concentration of
$3.9\times 10^{13}$ cm$^{-2}$, the surface electron states will
coexist with the bulk states and the band-bending induced 2DEG
states \cite{35} in Bi$_{\rm 2}$Se$_{\rm 3}$. The Fermi wavelength
in the bulk is estimated to be comparable to the thickness of the
nanoplates, so that all types of states will be influenced by
$V_{\rm g}$, not only on their concentration but also on spin
precession. If there are frequent scattering events between these
states, the tunability measured should represent an averaged value
over all the states. Otherwise, the type of states with higher
tunability was detected.

To summarize, Bi$_{\rm 2}$Se$_{\rm 3}$ is a very promising material
in terms of electrical control of spin dynamics. The appearance of
AB, AAS and AC quantum interferences in Bi$_{\rm 2}$Se$_{\rm 3}$
ring structures would allow more sophisticated multiple-degree
controls of spins in specially designed devices, to realize novel
functionalities such as possible non-Abelian operations on spins.

\begin{acknowledgments}
We would like to thank X. C. Xie, Q. F. Sun, Zhong Fang, Xi Dai and
T. Xiang for stimulative discussions, and H. F. Yang for
experimental assistance. This work was supported by NSFC, the
National Basic Research Program of China from the MOST under the
contract No. 2011CB921702, the Knowledge Innovation Project and the
Instrument Developing Project of CAS.

\end{acknowledgments}

\end{document}